\documentclass[letter,structabstract]{aa}  
\usepackage{graphicx}
\usepackage{natbib}
\bibpunct{(}{)}{;}{a}{}{,} 
\usepackage{txfonts}
\begin{document}
   \title{The nucleus of 103P/Hartley~2, target of the \textit{EPOXI} mission\thanks{Based on observations collected at the European Southern Observatory, Chile, under program 280.C-5077}}


   \author{C. Snodgrass
          \inst{1,2}
          \and
          K. Meech\inst{3,4}
          \and
          O. Hainaut\inst{5}
          }

\offprints{C. Snodgrass, \email{snodgrass@mps.mpg.de}}

\institute{Max Planck Institute for Solar System Research, Max-Planck-Strasse 2,
  37191 Katlenburg-Lindau, Germany
         \and
         European Southern Observatory, Alonso de C\'{o}rdova 3107,
  Vitacura, Casilla 19001, Santiago de Chile, Chile
  \and 
      Institute for Astronomy, 
             2680 Woodlawn Drive, Honolulu, HI 96822, USA
         \and
            NASA Astrobiology Institute
         \and
             European Southern Observatory,
             Karl-Schwarzschild-Strasse 2,
	     D-85748 Garching bei M\"unchen, Germany
            }

   \date{Received ; accepted }

 
  \abstract
   {103P/Hartley~2 was selected as the target comet for the \textit{Deep Impact} extended mission, \textit{EPOXI}, in October 2007. There have been no direct optical observations of the nucleus of this comet, as it has always been highly active when previously observed.}
   {We aimed to recover the comet near to aphelion, to a) confirm that it had not broken up and was in the predicted position, b) to provide astrometry and brightness information for mission planning, and c) to continue the characterisation of the nucleus.}
   {We observed the comet at heliocentric distances between 5.7 and 5.5 AU, using FORS2 at the VLT, at 4 epochs between May and July 2008. We performed $VRI$ photometry on deep stacked images to look for activity and measure the absolute magnitude and therefore estimate the size of the nucleus.}
   {We recovered the comet near the expected position, with a magnitude of $m_R = 23.74\pm0.06$ at the first epoch. The comet had no visible coma, although comparison of the profile with a stellar one showed that there was faint activity, or possibly a contribution to the flux from the dust trail from previous activity. This activity appears to fade at further epochs, implying that this is a continuation of activity past aphelion from the previous apparition rather than an early start to activity before the next perihelion. Our data imply a nucleus radius of $\le 1$ km for an assumed 4\% albedo; we \emph{estimate} a $\sim 6\%$ albedo. We measure a colour of $(V-R)=0. 26\pm0.09$.}
   {}

   \keywords{Comets: individual: 103P/Hartley 2}

   \maketitle
%

\section{Introduction}

The \textit{EPOXI} mission\footnote{\texttt{http://epoxi.umd.edu}} is a combination of two missions that reuse the surviving fly-by part of the \textit{Deep Impact} (DI) spacecraft. These missions were previously known as \textit{DIXI} (Deep Impact eXtended Investigation) and \textit{EPOCh} (Extrasolar Planet Observation and Characterisation), both of which were designed to use the DI high resolution camera, but with quite different scientific goals. The first was a direct extension of the original DI mission to Comet 9P/Tempel~1 \citep{Ahearn05}, aimed at furthering our understanding of the nuclei of Jupiter Family Comets (JFCs) by flying on to a second comet and imaging its nucleus in detail. The second part of the mission (EPOCh) makes use of the high resolution camera as a small aperture space telescope to study transiting extrasolar planets while the spacecraft is enroute to its comet target.

Following the DI encounter with 9P/Tempel~1 the spacecraft was left on an orbit that gave the possibility of redirecting it to one of two JFCs; 85P/Boethin or 103P/Hartley~2. Comet Boethin was the preferred target as it would mean a shorter mission, however it has not been observed since 1986 and could not be recovered despite a deep search over a large area of sky covering the entire orbital position error ellipse \citep{Meech10}. Therefore Comet Hartley~2 was selected as the target in October 2007, and the DI spacecraft was placed onto the trajectory that will take it to a flyby on November 4th 2010. 103P/Hartley~2 (hereafter, 103P) has the advantage that it has been regularly observed since it was discovered \citep{Hartley1986} and has a well determined orbit. The disadvantage of this target is that, at the time it was selected, its nucleus had never been directly observed.

Previous observations of 103P have always found the comet to be highly active, even at relatively large heliocentric distance. Ground based telescopic observations were consistent with a radius $r_{\rm N} \le 4$ km, but place only weak limits due to the high levels of activity observed \citep{Licandro00b,Lowry+Fitzsimmons01,Lowry03,Snodgrass06,Snodgrass08,Epifani08}. The previous observations are summarised in table \ref{prev_obs}. The previous best measurement of the nucleus radius comes from \citet{Groussin04}, who observed the comet when it was highly active at $R_{\rm h} = 1.2$ AU, using the \textit{Infrared Space Observatory} (ISO). They applied a coma subtraction technique previously used by this group with HST data \citep[e.g.][]{Lamy99}, where the high spatial resolution possible using space based (diffraction limited) observations of nearby comets allows separation of the nucleus signal from the surrounding coma. Using this technique, these authors found a nucleus radius of $0.71\pm0.13$ km for 103P. Interestingly, for this nucleus to produce the high level of activity seen in 103P, 100\% of the surface must be active, which goes against the general tendency of JFCs to have small fractional active areas \citep{Ahearn95}. \citet{Lisse09} obtained images of the comet using the \textit{Spitzer} space telescope. They also find a small nucleus with a large active area; their results are discussed further in section \ref{discussion}.

%
\begin{table}
\begin{minipage}[t]{\columnwidth}
\caption{Previous observations and nucleus size measurement for 103P.}
\label{prev_obs}
\centering
\renewcommand{\footnoterule}{}  
\begin{tabular}{ccccccl}
\hline \hline
Date & $R_{\rm h}$\footnote{Heliocentric distance (AU): All post-perihelion except for L09.} & $\Delta$\footnote{Geocentric distance (AU).} & $\alpha$\footnote{Solar phase angle (degrees).} & $m_R$\footnote{$R$-band magnitude.} & $r_{\rm N}$\footnote{Nucleus radius (or upper limit) in km assuming 4\% albedo and phase function of 0.035 mag deg$^{-1}$} & ref.\footnote{L00: \citet{Licandro00b}; LF01: \citet{Lowry+Fitzsimmons01}; L03: \citet{Lowry03}; S06, S08: \citet{Snodgrass06,Snodgrass08}; ME08: \citet{Epifani08}; G04: \citet{Groussin04}; L09: \citet{Lisse09}} \\
\hline
\multicolumn{6}{l}{Ground-based optical:}\\
28/03/93 & 4.7 & 3.8 & 3.3 & 20.1 & $\le 5.3$ & L00 \\
10/12/98 & 3.6 & 3.9 & 14.3 & 19.0 & $\le 6.4$ & LF01 \\
08/06/99 & 4.6 & 4.2 & 12.5 & 20.1 & $\le 6.3$ & L03 \\
06/03/05 & 3.2 & 2.3 & 5.0 & 18.7 & $\le 4.1$ & S06 \\
03/04/05 & 3.4 & 2.5 & 7.9 & 19.1 & $\le 4.3$ & ME08 \\
01/03/06 & 5.0 & 4.3 & 8.3 & 21.0 & $\le 4.5$ & S08 \\
\hline
\multicolumn{6}{l}{Space-based infrared:}\\
05/02/98 & 1.2 & 0.9 & 53.2 & -- & $0.71\pm0.13$ & G04\footnote{Based on a coma subtraction technique.} \\
12/08/08 & 5.5 & 4.9 & 9.5 & -- & $0.57\pm0.08$ & L09\\
\hline
\end{tabular}
\end{minipage}
\end{table}


\section{Observations}

Following the selection of 103P as the \textit{EPOXI} target, we were awarded Director's Discretionary time to directly observe the nucleus using the VLT+FORS2. We sought to apply the same technique that we have previously applied to the study of JFC nuclei; observation when the comet is at large heliocentric distance and therefore more likely to be inactive \citep{Meech04,Snodgrass05,Snodgrass06,Snodgrass08}. As noted above, 103P has shown considerable activity far from perihelion, so we observed it as near to aphelion as possible. 103P passed through aphelion at 5.88 AU on 7th August 2007. The first opportunity to observe it after it was selected as the \textit{EPOXI} target was April 2008. We obtained 4 epochs of observation starting in early May 2008 when the comet was still at $R_{\rm h} \ge 5.5$ AU; see table \ref{VLT_obs} for a VLT observation log. 

%
\begin{table*}
\begin{minipage}{\textwidth}
\caption{2008 Observations taken with VLT+FORS2.}
\label{VLT_obs}
\centering
\renewcommand{\footnoterule}{}  
\begin{tabular}{ccccccccccc}
\hline \hline
Date & $R_{\rm h}$ & $\Delta$ & $\alpha$ & Filter & $t_{\rm exp}$ & $N_{\rm exp}$\footnote{The number of exposures, with the number used in the stacked images (i.e. those clear of background stars) in brackets.} & $m_{\rm apparent}$ & $m_{\rm coma}$\footnote{Coma magnitude for a steady state coma, estimated from surface brightness profile \citep[see][]{Snodgrass05}} & $m_{\rm absolute}$ & $r_{\rm N}$\\
\hline
May 05.2 & 5.7 & 4.7 & 1.6 & $R$ & 74 & 43(23) & $23.74\pm0.06$ & $25.1\pm0.3$ & $16.55\pm0.06$ & $\le 1.40\pm0.03$ \\
May 05.3 & 5.7 & 4.7 & 1.6 & $V$ & 74 & 43(34) & $24.01\pm0.07$ & -- & $16.81\pm0.07$ & $\le 1.46\pm0.05$ \\
May 06.2 & 5.7 & 4.7 & 1.4 & $I$ & 74 & 43(37) & $24.31\pm0.10$ & $27.1\pm1.0$ & $17.12\pm0.10$ & $\le 0.92\pm0.04$ \\
Jun 01.2 & 5.6 & 4.7 & 4.0 & $R$ & 100 & 24(16) & $23.69\pm 0.09$ & -- & $17.30\pm0.09$ & $\le 0.99\pm0.04$ \\
Jun 04.2 & 5.6 & 4.7 & 4.5 & $R$ & 83 & 28(28) & $24.50\pm 0.04$ & $25.4\pm0.2$ & $17.25\pm0.04$ & $\le 1.01\pm0.02$ \\
Jul 28.0 & 5.5 & 5.2 & 10.5 & $R$ & 240 & 11(11)\footnote{The comet was in front of background stars throughout these observations; see section \ref{sec:july}. The quoted uncertainty on the photometry does not include the estimated $\pm0.2$ mag. uncertainty due to the background subtraction process.} & $25.06\pm 0.12$ & $26.2\pm0.2$ & $17.41\pm0.12$ & $\le 0.94\pm0.06$ \\
\hline
\end{tabular}
\end{minipage}
\end{table*}

All data were taken in service mode at the VLT using the direct imaging mode of the FORS2 camera on UT1, with broad-band $VRI$ filters. Frames were tracked at the sidereal rate, with exposure times kept short enough to keep the comet's motion $\le 0.5\arcsec$, i.e. within the seeing disk (0.5--1.1\arcsec{} FWHM for these runs, median 0.7\arcsec). This method has the advantage that the image PSF can be accurately reconstructed from the field stars to compare with the comet profile. The basic data reduction (bias subtraction, flat fielding) was done in the normal way, using IRAF.  

When the comet was detectable in individual frames we performed photometry on each (rejecting those where the comet was too close to field stars), measuring the brightness relative to field stars. We used an aperture radius matched to the seeing to maximise $S/N$, and an aperture correction and the calibrated magnitudes of the field stars to give accurate photometry within a standard (10\arcsec{} diameter) aperture. This method is described in more detail by \citet{Snodgrass05}. Unfortunately insufficient standard stars were observed on the night of May 5, but we were able to obtain calibrated magnitudes for the field stars by observing the field on a later photometric night. At the epochs where we had to stack the images to recover the comet the photometry was performed on the combined image. This photometry was calibrated directly using standard stars observed on the same nights.


\section{Results}
\label{results}

\subsection{May 2008}

   \begin{figure}
   \centering
   \begin{tabular}{l r}
   \raisebox{0.7cm}{\includegraphics[width=0.19\textwidth]{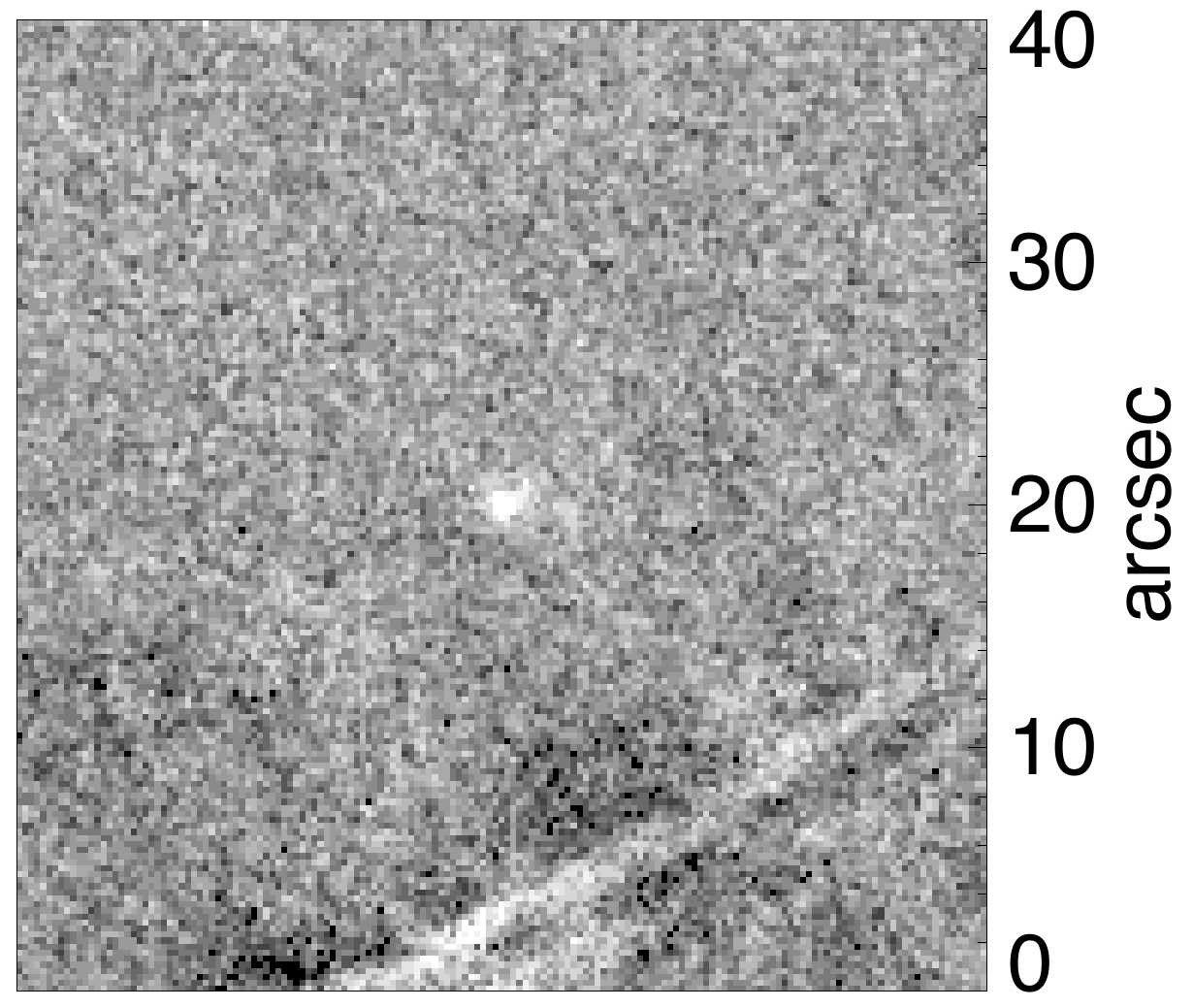}} &
   \includegraphics[width=0.28\textwidth]{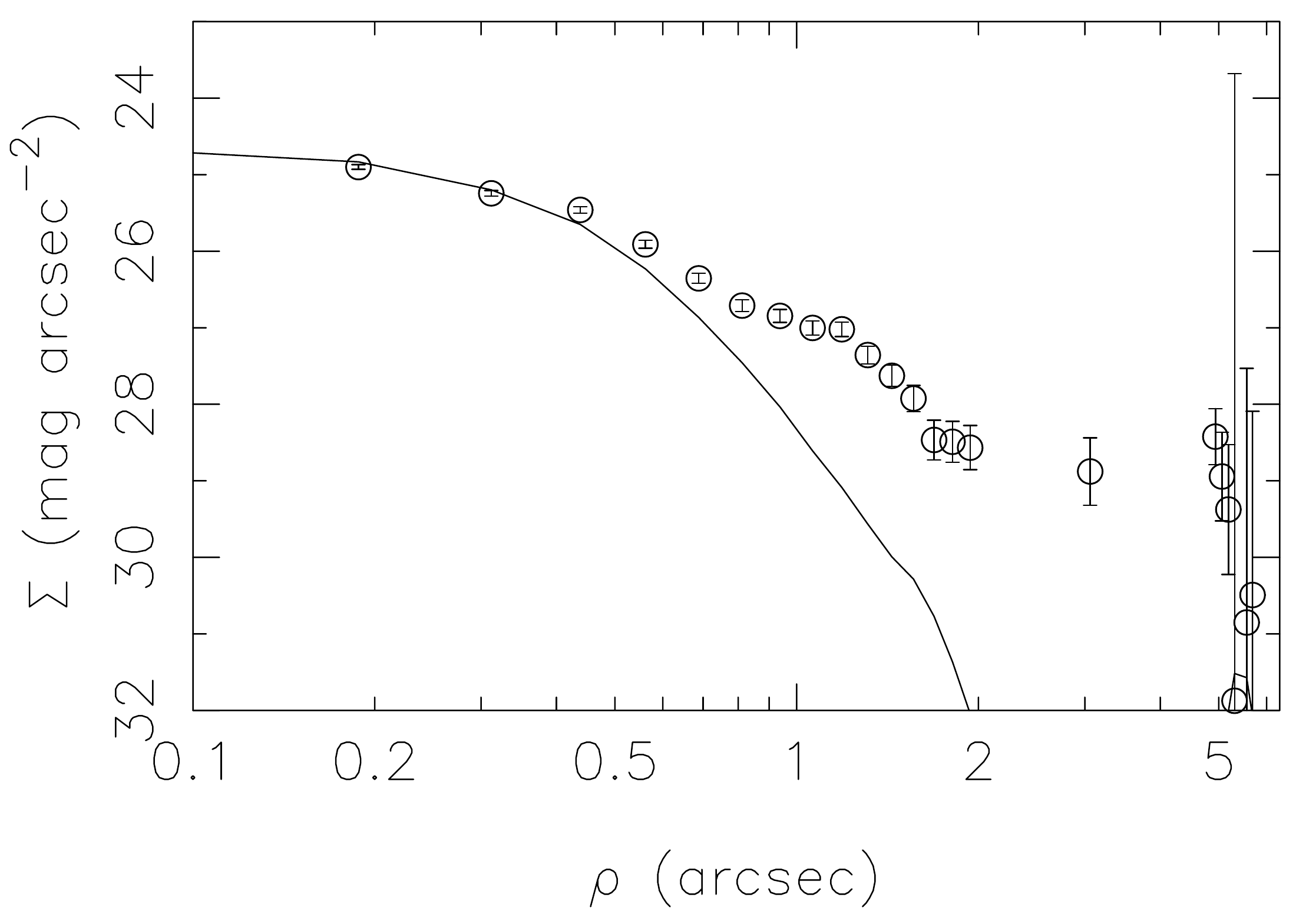} \\
   \end{tabular}
      \caption{$R$-band image and surface brightness profile taken on 2008 May 05.2 UT. The image is a stacked image ($23 \times 74$ s)  with background subtraction applied. The features to the lower right of the image are residuals from saturated stars that could not be entirely removed. The profile compares the comet (points) with the stellar PSF (solid line) measured in the equivalent stacked image of the star field.}
         \label{May_R}
   \end{figure}
%
   \begin{figure}
   \centering
   \begin{tabular}{l r}
   \raisebox{0.65cm}{\includegraphics[width=0.19\textwidth]{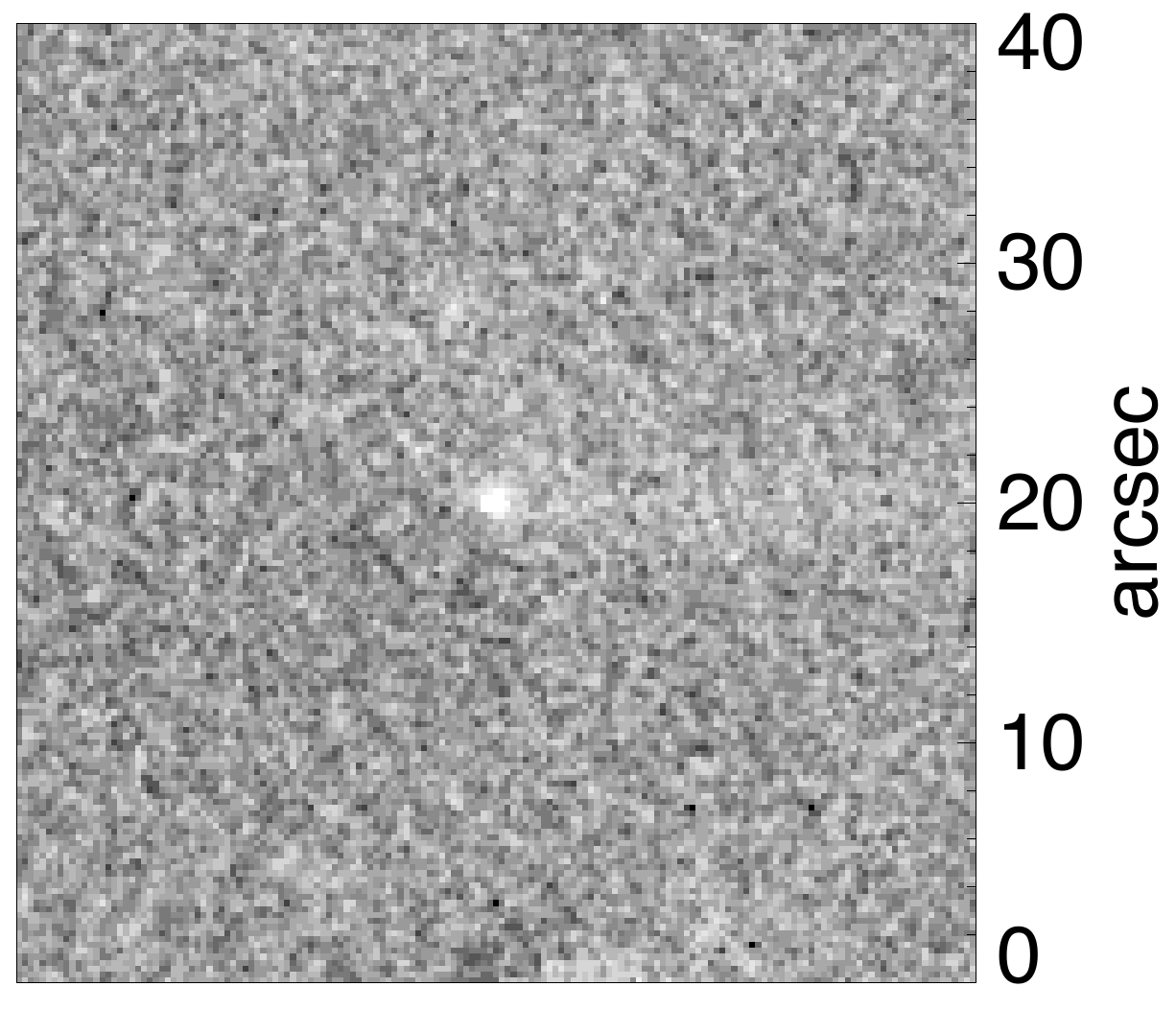}} &
   \includegraphics[width=0.28\textwidth]{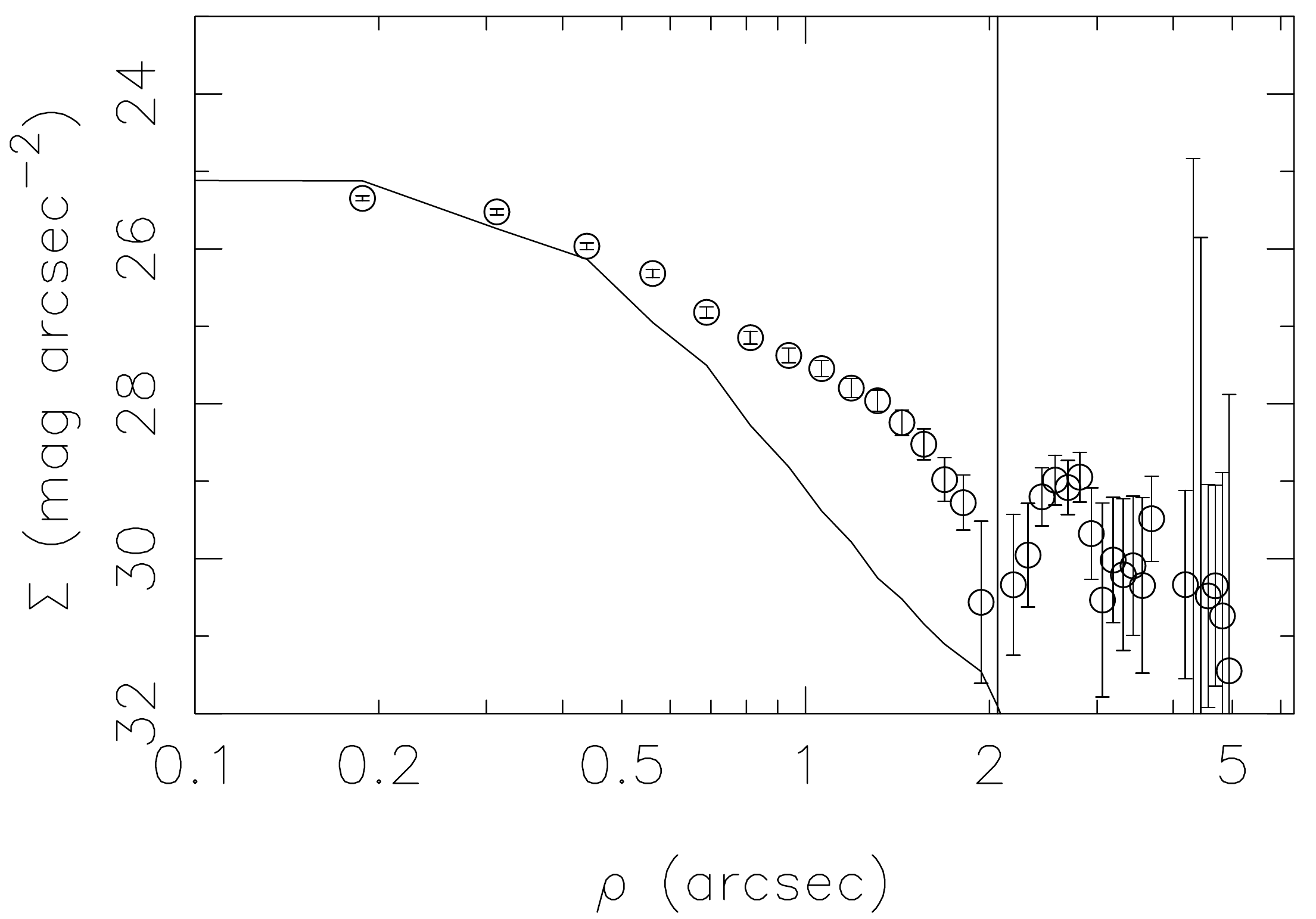} \\
   \end{tabular}
      \caption{$I$-band image ($37 \times 74$ s) and profile taken on 2008 May 06.2 UT. A trail is apparent in the image to the West (right) of the nucleus.}
         \label{May_I}
   \end{figure}

We were able to recover the comet near the nominal position in individual frames in the $R$-band (and marginally in the $V$). The individual frames were then stacked based on the comet's motion to provide a deep image (here we rejected frames in which the comet passed near to background stars). In the stacked images the comet was clearly visible in all filters, and had no visible coma or tail (Fig. \ref{May_R}). We searched for any faint activity around the comet by comparing the radial profile measured in the stacked $R$-band image with the stellar PSF (measured in an equivalent stacked image of the background star field to take into account the effect of any variation in seeing between the images). The surface brightness profile suggests that there is faint coma around the nucleus even at $R_{\rm h}=5.7$ AU post-aphelion (Fig. \ref{May_R}), as there is excess flux above the PSF.  It is also possible that the excess flux is entirely due to reflections from grains in the trail of material left from previous active periods that fall within the aperture due to projection effects. The roughly linear part of the profile at around $\rho = 1\arcsec$ has a gradient near -1.5, as expected for steady state coma affected by radiation pressure \citep{Jewitt+Meech87}, however the noisy background due to residuals from the nearby saturated stars makes it difficult to draw any strong conclusions from the profile shape. In either case, the `nucleus' magnitudes measured therefore only give limits on the true absolute magnitude and radius of the nucleus.

We measure the magnitude of the nucleus to be $m_R = 23.74\pm0.06$, which gives an absolute magnitude of $m_R(1,1,0) \ge 16.55$ assuming a linear phase coefficient of $\beta = 0.035$ mag deg$^{-1}$. This implies a radius of $r_{\rm N} \le 1.40\pm0.03$ km (assuming a typical 4\% albedo), which is larger than found by \citet{Groussin04} and \citet{Lisse09} but a significantly stronger limit than obtained from other optical observations.

There is no significant variation in individual magnitudes in either of the 1.2 hour long $R$- or $V$-band sequences; this implies either a long rotation period, a near spherical nucleus, a pole on orientation or the coma masking the modulation due to the rotation of the nucleus, 
or some combination of these effects. A longer dedicated sequence of observations will be required to constrain the shape and rotational state of the nucleus.

As the $V$- and $R$-band sequences were taken consecutively we can use them to measure the colour of the comet. The $V$-band magnitude was measured to be $24.01\pm0.07$. This gives $(V-R) = 0.26\pm0.09$, which is very blue for a JFC nucleus [JFC mean $(V-R) = 0.45\pm0.11$; \citet{Snodgrass08}], but the flux measured here is not 100\% from the nucleus. It is consistent with the reported comae colours measured for 103P by \citet{Lowry03} and \citet{Snodgrass08} who found $(V-R)=0.32\pm0.12$ and $(V-R)=0.16\pm0.09$ respectively when the comet was much more active. \citet{Licandro00b} measured a redder colour of $(V-R)=0.5\pm0.1$ for the coma, but this value is still consistent at 2$\sigma$ with our result.

We also measured $m_I = 24.31 \pm 0.10$, although we make no attempt to measure a $(R-I)$ colour as the $I$-band observations were performed one night later than the $V$- and $R$-band. \citet{Snodgrass08} measured coma colours of $(R-I)=0.35\pm0.08$ when the comet was outbound at $R_{\rm h} = 5.0$ AU. 
On May 6 the comet was further from bright background stars, and the cleaner background subtraction allows us to detect a trail along the plane of the orbit (at position angle $\sim 270\degr$) in the stacked image (Fig. \ref{May_I}). The average surface brightness of the trail was measured to be $27.8 \pm 0.1$ mag arcsec$^{-2}$ over an area of 21 pixels (1.3 arcsec$^2$) around 10\arcsec{} from the nucleus in this image. The surface brightness profile has a similar shape to the one measured the previous night in the $R$-band, with a reasonably linear slope around $\rho = 1\arcsec$, and again is indistinguishable from the noise in the sky beyond 2\arcsec.

\subsection{June 2008}

   \begin{figure}
   \centering
   \begin{tabular}{l r}
   \raisebox{0.71cm}{\includegraphics[width=0.19\textwidth]{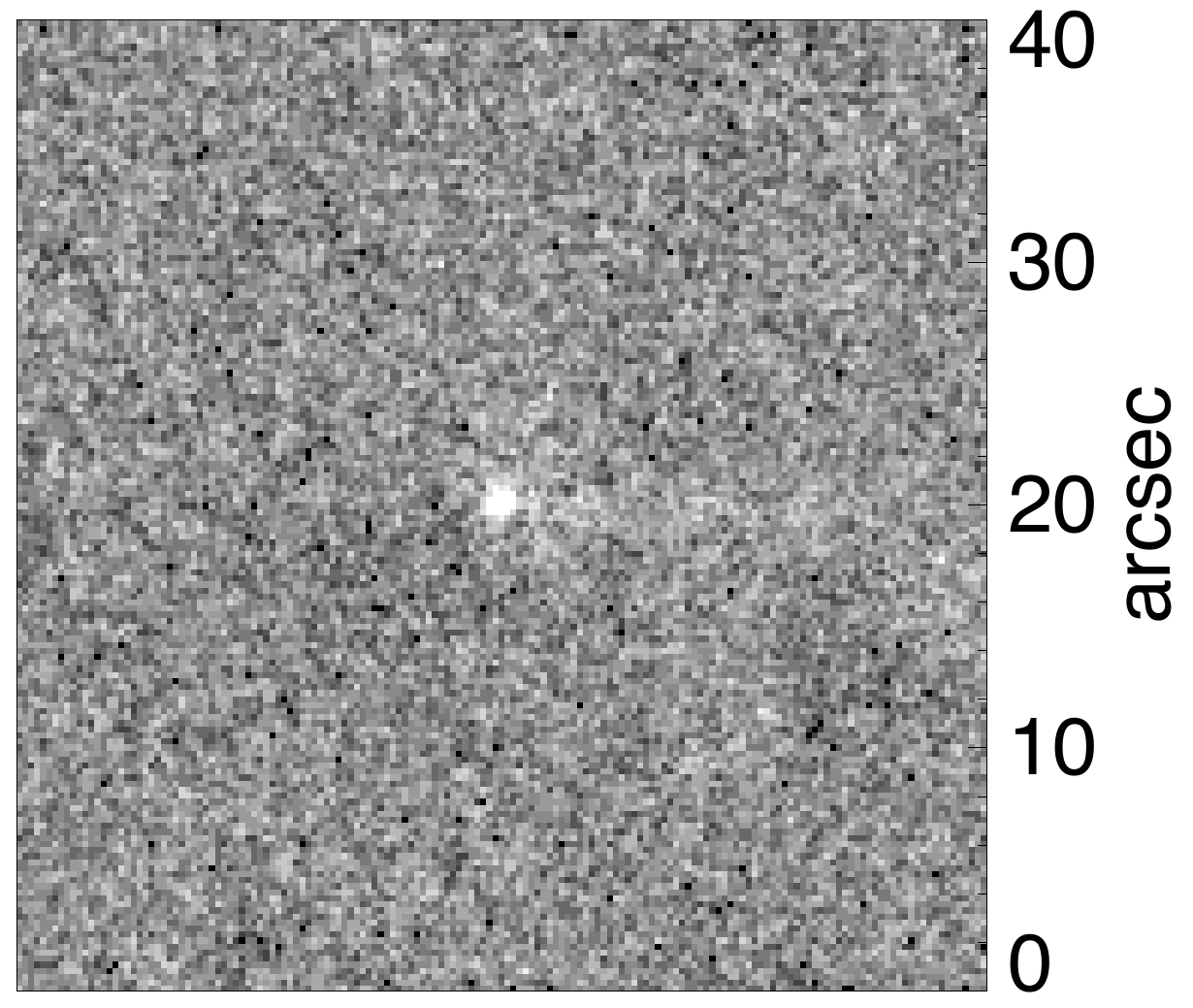}} &
   \includegraphics[width=0.28\textwidth]{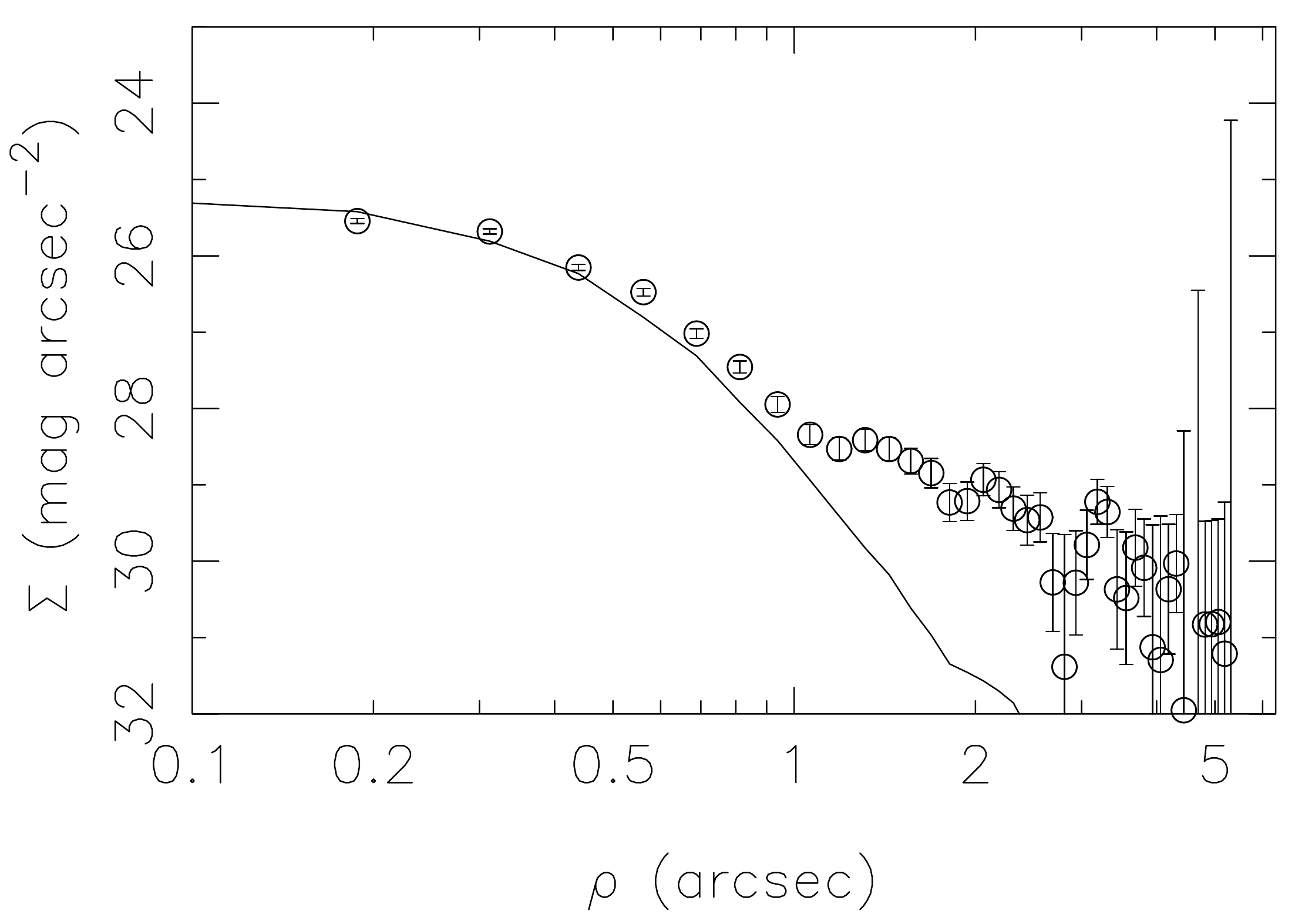} \\
   \end{tabular}
      \caption{$R$-band image ($28 \times 83$ s) and profile taken on 2008 June 04.2 UT. A trail is apparent in the image to the West (right) of the nucleus.}
         \label{June_R}
   \end{figure}
%

   \begin{figure}
   \centering
   \begin{tabular}{l r}
   \includegraphics[width=0.23\textwidth]{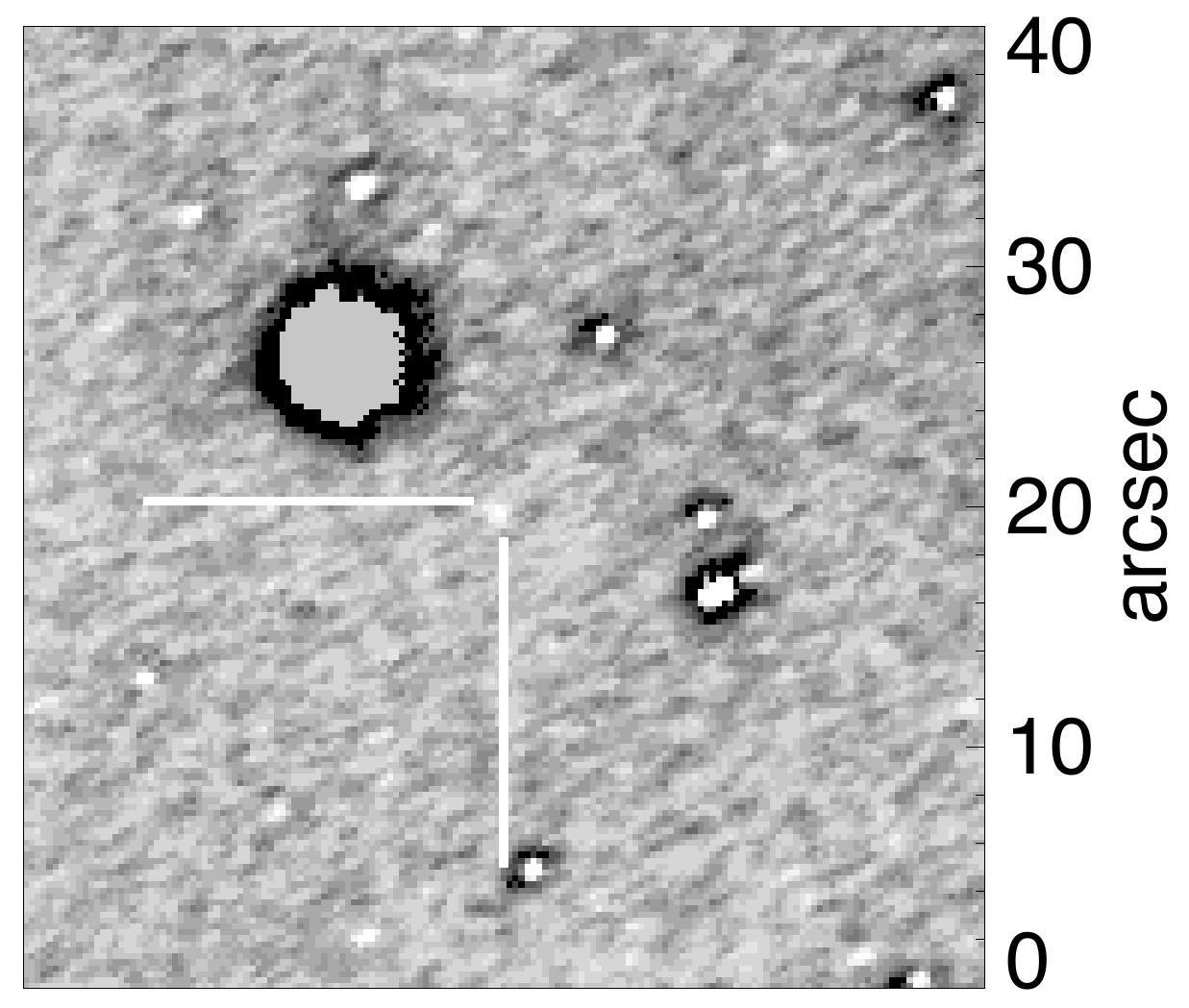} &
  \includegraphics[width=0.23\textwidth]{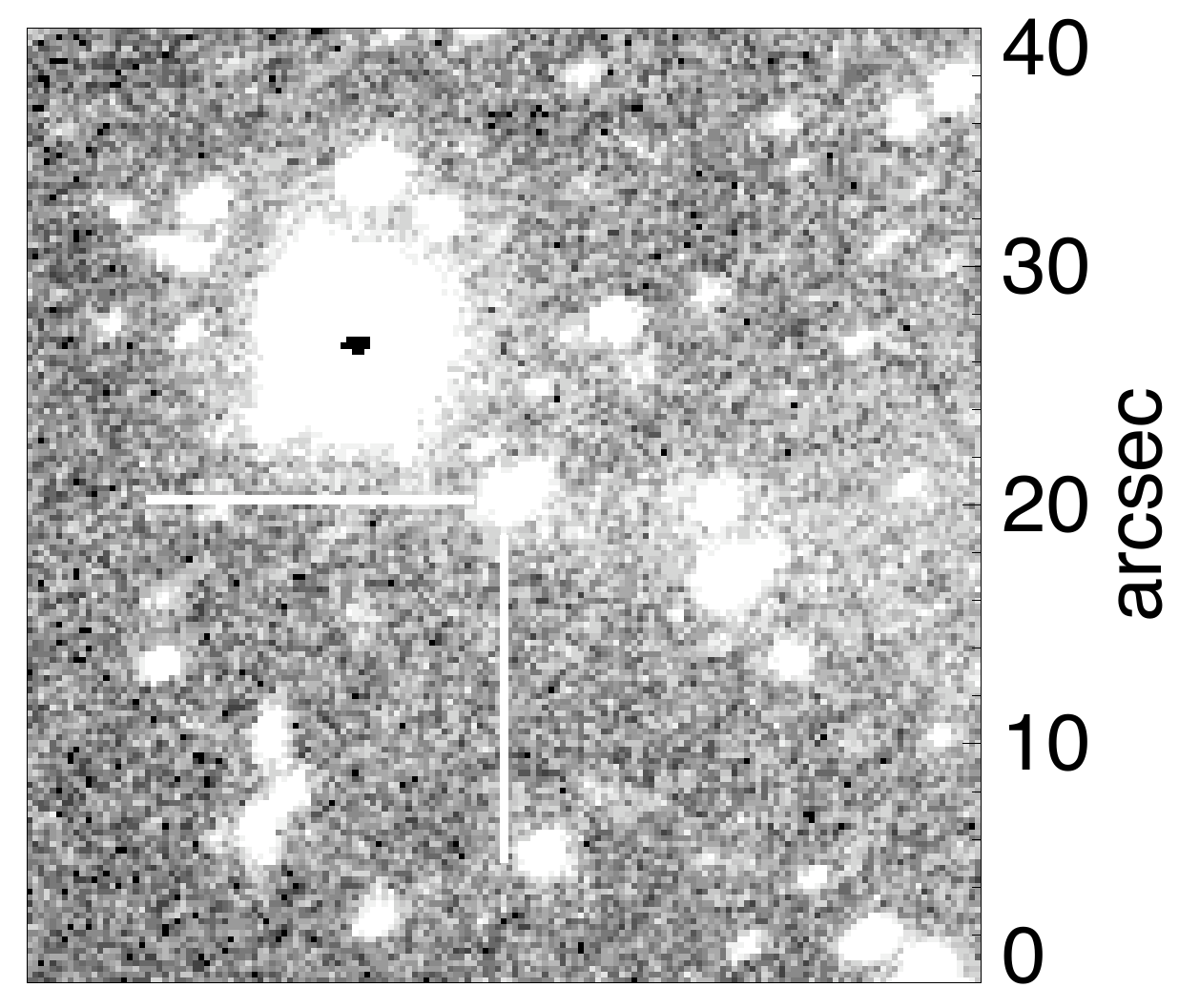} \\
   \end{tabular}
      \caption{$R$-band image ($11 \times 240$ s) and taken on 2008 July 28.0 UT, with and without background subtraction applied.}
         \label{July_R}
   \end{figure}

103P was observed at 3 further epochs at the VLT, but in only the $R$-band. On June 1.2 UT the conditions were not ideal and the comet was found to be very close to a bright field star, making photometry impossible in individual frames. A background subtraction routine was used to leave an image of the comet in the combined frame, in which we measure $m_R=24.54\pm0.07$. The uncertainty on this value includes photometric and calibration uncertainties, but not any additional uncertainty due to possible incomplete conservation of the comet flux in the background subtraction. The background subtraction leaves some negative residuals due to the fact that the nearby star saturated and therefore perfect scaling of the reference star field image was not possible.  This also made it impossible to measure a reliable surface brightness profile.

The comet was observed in a clear area of sky on June 4.2 UT, only passing near to two very faint background stars. We were able to measure a clean profile in a combined background subtracted image (Fig.~\ref{June_R}) for assessing any potential activity. The profile is a closer match to the stellar PSF in the inner region than the May one, possibly implying a decrease in activity during this period. The dust trail is just visible in this image, with a surface brightness of $\sim 28.5$ mag arcsec$^{-2}$.

Individual magnitudes show a slight trend upwards towards the end of the sequence, although this is also when the comet approaches the faint stars, so is not a reliable indication of photometric variation in the comet. The mean magnitude is found to be $m_R=24.50\pm0.04$. 

\subsection{July 2008}
\label{sec:july}

On July 28.0 the comet was, by chance, directly in front of two faint field stars (or possibly a galaxy), and moving so slowly on the sky so that it remained so throughout the service mode observation. As we could not generate a background image from the data itself due to the small motion of the comet we took an image of the same star field at a later date to subtract from the data. Unfortunately the seeing in this background image was significantly worse than in the original data and the sky level was very high, making a clean background subtraction impossible: Figure~\ref{July_R} shows residual `donuts' around brighter stars and considerable noise in the sky. We measure $m_R = 25.06 \pm 0.12$ for the comet, including statistical and calibration uncertainties. We assess the contribution to the uncertainty due to the imperfect background subtraction by adding fake `comets' with the same rate of motion to the individual frames before subtraction, and testing our ability to recover them in both regions of clear sky and when placing them on top of other faint background sources. We find that the differences between the input and measured magnitudes of our fake comets are between 0.1 and 0.4 magnitudes, and take $\pm 0.2$ magnitudes as a reasonable assessment of the average uncertainty in the photometry introduced by the background subtraction process. This dominates over the other sources of statistical uncertainty, however there remains the possibility that there is a systematic over or under correction at the position of the comet.

In Fig.~\ref{July_R} we also show a combined image without any background subtraction, in which the very small motion of the comet is apparent from the short length of the star trails. In this image the trail from the comet remains just detectable to the eye, although it is of such low surface brightness that it is entirely lost in the sky noise in the background subtracted image.


\section{Discussion}
\label{discussion}

The comet is seen to fade during the 3 months of observation, despite moving closer to the Sun during this period. The $R$-band magnitudes reduced to unit heliocentric and geocentric distance and zero phase angle (assuming a linear phase function of 0.035 mag deg$^{-1}$) are $m_R(1,1,0)=16.55$, 17.30, 17.25 and 17.41 for the 4 epochs between $R_{\rm h} = 5.7$ and 5.5 AU. At the level of uncertainty of our measurements the last 3 all agree, but the first epoch is significantly brighter.
There appears to be near nucleus activity in the first data, and less in the June 4th data. Together, these results suggest that 103P remains active past aphelion, but that this activity decreases until the comet probably reaches an inactive state before the next season of activity starts pre-perihelion. This is in agreement with the \textit{Spitzer} observations of \citet{Lisse09} who found the comet to be inactive at 5.5 AU in August 2008, $\sim 2$ weeks after our last epoch. \citet{Lisse09} also detect the dust trail that we see in some of our data; they conclude that it consists of mm-sized particles from the last (May 2004) perihelion passage.

The nucleus radii implied by these magnitudes (for an assumed albedo of 4\%) are $r_{\rm N} \le 1.4$, 1.0 and 0.9 km for the May, June and July epochs respectively. These limits are much stronger than previous observations of the comet when active, as it is clearly only weakly active in these exposures. The radius found by \citet{Lisse09} is $0.57\pm0.08$ km, and these authors used a preliminary value for our July $R$-band magnitude of $m_R(1,1,0)=18.9$ (obtained before we were able to take the background frame for subtraction, it most likely \emph{underestimates} the brightness of the comet) to derive an albedo of 0.028. Using $m_R(1,1,0)=17.4\pm0.2$ instead we find an albedo of $0.11\pm0.02$, which would be very high for a cometary nucleus. The upper limit on the albedo from the June photometry, which has no uncertainty from background subtraction but includes flux from the coma or dust trail, is $\le 12\%$. 
If we disregard our uncertain July photometry and instead assume that the absolute magnitude continued to follow the increasing trend set in May and June (due to decreasing activity) then it would be $\sim 18$ at the time of the \textit{Spitzer} observations, implying a $\sim 6\%$ albedo for $r _{\rm N}=0.57$. 

\section{Conclusions}

We observed comet 103P/Hartley~2 using FORS2 at the VLT on 4 epochs while the comet was beyond 5.5 AU from the Sun, post-aphelion. We find:

   \begin{enumerate}
      \item The comet was not visibly active in combined images at each epoch, however comparison of the profile with the image PSF revealed that there was some near nucleus dust present, even at this large distance pre-perihelion. The level of activity appeared to decrease between May and June as the comet moved inbound, implying that this was residual activity from the last active period, not the start of activity before the next perihelion passage, although we cannot rule out that this represents the trail of dust from previous perihelion passages.
     \item The measured photometry puts stronger constraints on the nucleus magnitude than any previous optical observations as any activity was very weak. From the June data we find that the radius of the comet is $\le 1$ km (assuming a 4\% albedo). Using $r _{\rm N}=0.57$ \citep{Lisse09} we find that the albedo is $\le 12\%$.
     \item The brightness decrease, and the observation of an inactive nucleus by \citet{Lisse09} in August 2008, implies the activity ended by late July or early August, at $R_h = 5.5$ AU post-aphelion. Unfortunately our photometry from late July remains uncertain due to the comet flux being merged with background sources, but an extrapolation from May and June gives $m_R(1,1,0) \approx 18$ for the inactive nucleus and an albedo of $\sim 6\%$.
      \item Finally, we measure the colour of the nucleus to be $(V-R)=0.26\pm0.09$, which is bluer than found for JFC nuclei but is affected by the activity.
   \end{enumerate}

\begin{acknowledgements}
      We thank ESO for awarding Director's Discretionary Time to this project, the Paranal astronomers and telescope operators who performed the observations for us and the referee, Dr.~S.~C.~Lowry, for helpful comments that improved this paper. This work is based in part on support from the EPOXI mission through a subcontract from the University of Maryland Z631506.
\end{acknowledgements}

\bibliographystyle{aa} 
\bibliography{14790tex}  

\end{document}